# Room-temperature control and electrical readout of individual nitrogen-vacancy nuclear spins

**Michal Gulka**[1,2,3*†], **Daniel Wirtitsch**[4,5*], **Viktor Ivády**[6,7], **Jelle Vodnik**[1,2], **Jaroslav Hruby**[1,2], **Goele Magchiels**[1], **Emilie Bourgeois**[1,2], **Adam Gali**[6,8], **Michael Trupke**[4,5], **Milos Nesladek**[1,2,3†]

[1]Institute for Materials Research (IMO), Hasselt University, Wetenschapspark 1, B-3590 Diepenbeek, Belgium.
[2]IMOMEC division, IMEC, Wetenschapspark 1, B-3590 Diepenbeek, Belgium.
[3]Department of Biomedical Technology, Faculty of Biomedical Engineering, Czech Technical University in Prague, Sítná sq. 3105, 272 01, Kladno, Czech Republic.
[4]Faculty of Physics, University of Vienna, VCQ, Boltzmanngasse 5, 1090 Vienna, Austria.
[5]Institute for Atomic and Subatomic Physics, Vienna University of Technology, VCQ, Stadionallee 2, 1020 Vienna, Austria.
[6]Wigner Research Centre for Physics, PO. Box 49, Budapest H-1525, Hungary.
[7]Department of Physics, Chemistry and Biology, Linkoping University, SE-581 83 Linköping, Sweden
[8]Department of Atomic Physics, Budapest University of Technology and Economics, Budafoki út 8., H-1111, Budapest, Hungary
[*]These authors contributed equally to this work
[†]Email: gulka.michal@gmail.com, milos.nesladek@uhasselt.be

## Abstract

Nuclear spins in semiconductors are leading candidates for quantum technologies, including quantum computation, communication, and sensing. Nuclear spins in diamond are particularly attractive due to their extremely long coherence lifetime. With the nitrogen-vacancy (NV) centre, such nuclear qubits benefit from an auxiliary electronic qubit, which has enabled entanglement mediated by photonic links. The transport of quantum information by the electron itself, via controlled transfer to an adjacent centre or via the dipolar interaction, would enable even faster and smaller processors, but optical readout of arrays of such nodes presents daunting challenges due to the required sub-diffraction inter-site distances. Here, we demonstrate the electrical readout of a basic unit of such systems – a single $^{14}$N nuclear spin coupled to the NV electron. Our results provide the key ingredients for quantum gate operations and electrical readout of nuclear qubit registers, in a manner compatible with nanoscale electrode structures. This demonstration is therefore a milestone towards large-scale diamond quantum devices with semiconductor scalability.

## Introduction

Here we introduce a novel approach for a quantum computation platform operating at room temperature via electrically-read electron-nuclear single-spin gates. This approach might significantly reduce technological hurdles for the realisation of future diamond quantum computing concepts. In recent works[1-6] we have developed a technique, in which the spin state of a coherently controlled negatively charged nitrogen-vacancy (NV) electron spin is read electrically by photoelectric detection of magnetic resonance (PDMR). Unlike the case of optical readout, for which spatial resolution is usually diffraction-limited, by employing PDMR, achievable spatial resolution for individual NVs is determined solely by the electrode size. The reduction in the interrogation area would allow for an individual readout of NV spins placed at the dipole-dipole interaction scale[7], which would be a significant step towards deterministically entangled qubit arrays. PDMR is based on the spin state dependent photo-ionization of NV, mediated by the charge state conversion[5]. The application of an external electric field accelerates the photo-generated charge carriers towards the electrodes fabricated on the diamond surface. Thus, the NV spin state can be determined by directly measuring the photocurrent without the need for photon collection, even for single NV centres[4]. PDMR features several additional benefits for spin readout since, in contrast to conventional optical detection schemes, it does not suffer from the NV excited state saturation behaviour at high incoming photon fluxes. In the case of photocurrent, the saturation is given instead by the charge carrier recombination lifetime in diamond, which is several orders of magnitude higher than the excited state lifetime[4]. In combination with the high electron collection efficiency, PDMR readout may therefore enable shot noise reduction due to significantly higher electron detection rates compared to photon counting[5].

To advance towards nanoscale NV qubit systems, we demonstrate electrical readout of its basic element, a two-qubit electron-nuclear spin system. Although the readout of ensembles of nuclear spins has recently been achieved by using electrically detected electron-nuclear double resonance (EDENDOR)[8], single NV nuclear spin electrical readout has remained elusive due to the long sequences required for nuclear spin manipulation (a factor of about 1000 compared to electron spin manipulation), resulting in a decrease in photocurrent and an excessive amount of DC noise accumulation during this period. In this work, we overcome these challenges and achieve room temperature photoelectric readout of the single intrinsic $^{14}$N nuclear spin of the NV centre, mediated by the electron spin at the excited-state level anti-crossing (ESLAC).

We show that by using a lock-in envelope detection technique, we are able to detect nuclear magnetic resonance spectra and Rabi oscillations with high contrast, even for the long intervals between laser excitation pulses required for nuclear spin manipulation. We also provide evidence that the ESLAC conditions allow for highly efficient MW-free protocols for the electrical detection of the nuclear spin. We further develop a theoretical model to describe photoelectric detection rates measured experimentally at the ESLAC, using the Lindblad master equation, which includes the charge-state transitions, time-dependent spin polarisation, and charge carrier readout. Future implementations of more complex schemes, such as the inclusion of nearby $^{13}$C nuclei, can now be envisaged aiming towards scalable quantum hardware for quantum computation and sensing.

## Results

1) Single NV photoelectrical detection and imaging

Throughout the experiments a commercial IIa high-pressure high-temperature (HPHT) diamond (New Diamond Technology, < 10 ppb background nitrogen) with intrinsic single NV defects was used. To collect charge carriers under bias voltage, coplanar interdigitated contacts with a 3.5 µm gap were fabricated on the diamond surface by means of optical lithography. The photocurrent was pre-amplified and then recorded via lock-in detection[3]. We stochastically allocated a few individual NV centres between the electrodes and selected those approximately 2.5 µm below the diamond surface. To find optimal qubit operational parameters, we first measured the DC current-voltage (I-V) characteristics, in which we identified three main current components (see Figure 1a): i) *NV photocurrent* – resulting from two-photon ionization of the single NV centre, ii) *non-NV photocurrent* – produced by photoionization of other diamond defects and iii) *dark current* - present after application of the voltage between electrodes without laser illumination. The sum of the dark and non-NV currents, denoted further as a background current, was determined from measurements with the laser beam focused between contacts, but off the NV centre. The measured DC I-V characteristics (see Figure 1b) show nonlinear characteristics, as discussed previously[4]. After having calculated the signal-to-background contrast (SBC), we set the bias voltage to the optimal 8.6 V, corresponding to the highest measured NV SBC (>65%). However, in some practical applications, an increase in bias voltage might prove beneficial to increase signal acquisition at the cost of SBC.

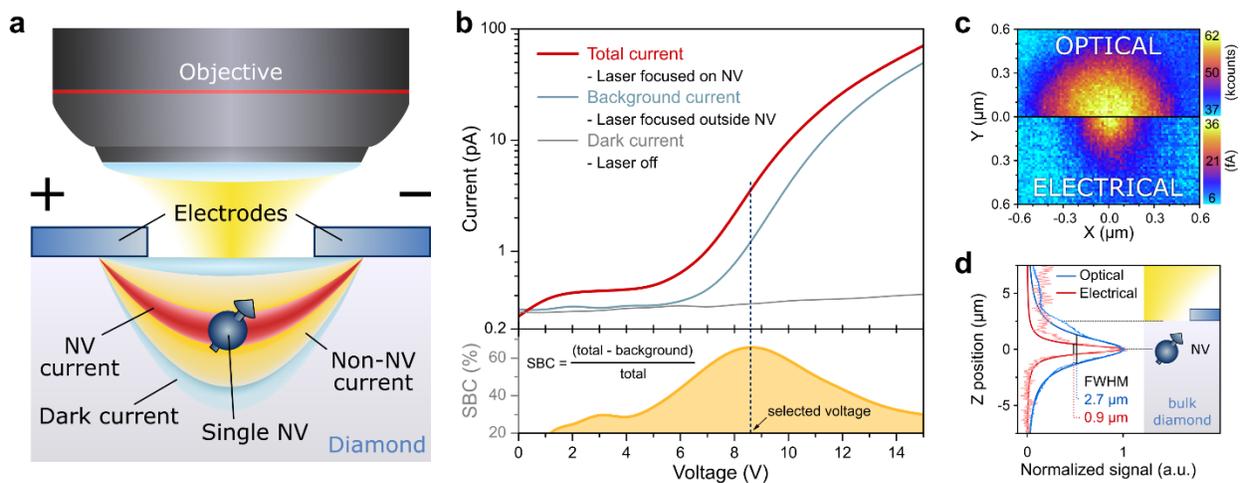

**Figure 1 | Electrical detection of single NV centre. a**, A schematic of the single NV centre PDMR chip used for the measurements. A yellow-green 561 nm laser is focused between the contacts (with an inter-electrode

distance of 3.5 µm) using a microscope objective in air (N.A. = 0.95). The resulting currents are measured versus the bias voltage applied to the electrodes. We identify three types of currents: *dark current* – not related to the laser illumination, *non-NV photocurrent* – laser-induced photocurrent not originating from the NV centre, *NV photocurrent* – current from the two-photon ionization of the single NV centre. **b**, Current-voltage characteristic curves for laser off (grey, *dark current*), for laser (4mW) focused away from the single NV centre (cyan, background current from *dark current* and *non-NV photocurrent*) and for laser (4mW) focused on the single NV centre (red, total current from *dark current*, *non-NV photocurrent*, and *NV photocurrent*). The bias voltage for the photocurrent measurements was set to 8.6 V as determined from the maximum signal-to-background contrast (**SBC**) calculated from the background and total current (dark yellow). **c-d**, Simultaneous optical and electrical imaging of the single NV centre (laser power 6 mW). **c**, XY map showing the size comparison of the same NV centre for the two detection methods. **d**, Z scan of the NV. Darker curves are the Lorentzian fits of the experimental data points.

In order to reduce the photocurrent contribution induced by photoionization of substitutional nitrogen (P1 centres, photoionization onset ~ 550 nm)[9,10] frequently present in diamond crystals, we used a yellow-green 561 nm (2.21 eV) laser instead of the commonly applied green excitation (532 nm). Optimal laser powers for photoelectric imaging of the single NV centre have been previously identified to be between 2 and 4 mW[4]. In order to enable detection of low average photocurrents imposed by the prolonged pulse sequences needed for nuclear spin manipulation (discussed below), higher laser powers (4 - 6 mW) are applied here to increase the NV-generated photocurrent. Even at these rather high laser powers, unlike photoluminescence, photocurrent does not saturate and a higher S/N ratio is reached. However, further increase of laser power would lead to a reduction of the NV magnetic resonance contrast unless the laser readout pulses are made correspondingly shorter[5]. The selected NV centre was imaged both electrically and optically in order to allow for a direct comparison between the two readout methods (see Figure 1c). Due to the two-photon nature of NV ionization and the Gaussian beam shape after the focussing objective, electrical imaging has been shown to improve spatial resolution and imaging contrast[4]. Here, we observe an even more substantial improvement in resolution in all three dimensions, with a threefold reduction of the axial size for electrical imaging (see Figure 1d).

2) Photoelectric readout principle at ESLAC

A schematic of the model describing the photoelectric readout principle at the excited-state level anti-crossing (ESLAC) is depicted in Figure 2a, including the charge transitions of the NV centre. The full photoionization model is discussed more in detail in Section 3 Modelling. To achieve photoelectric readout of a single nuclear spin using the NV centre electron spin as an ancilla, we first polarize the $^{14}$N nuclear spin to the $|m_I\rangle = |+1\rangle$ state[11–13]. When an external magnetic field of ~ 510 G (approaching the ESLAC) is aligned with the NV axis, the optical pumping polarizes the nuclear spin into the $|m_I\rangle = |+1\rangle$ state (spin polarization of >98%[11]) due to the spin state mixing and simultaneously initializes the electron spin into $|m_s\rangle = |0\rangle$[11–13]. This allows us to directly perform coherent nuclear spin control and sensitively read the spin state via photocurrent measurements, as the electron photoionization probability depends on the electron-nuclear spin states. In order to electrically detect nuclear hyperfine interactions, a pulsed lock-in envelope readout technique[3] was employed. In this detection method, readout laser pulses are solely applied during the high-state of the low-frequency envelope duty cycle, whilst the pulsed microwave (MW) driving is performed continuously. The rising edge of the frequency envelope is then used to trigger the lock-in amplifier. As can be seen in Figure 2b, PDMR clearly resolves the hyperfine spectrum of a single nuclear spin close to the ESLAC.

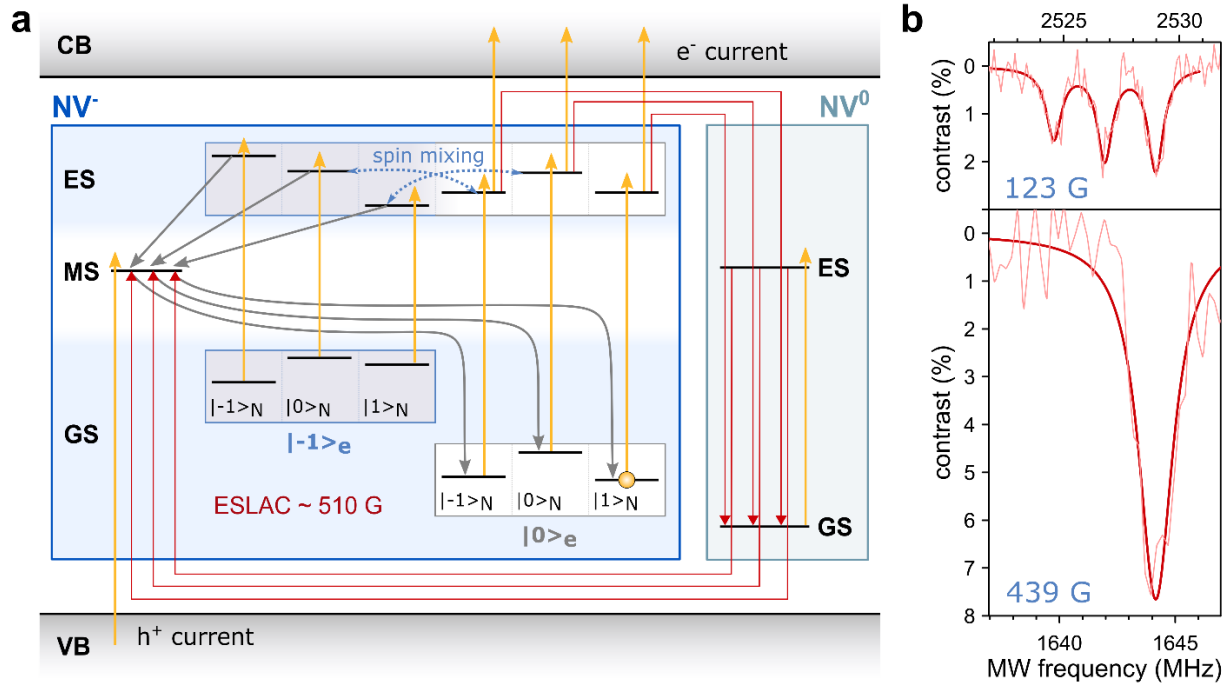

**Figure 2 | PDMR detection at ESLAC**. **a**, Schematic describing the photoelectric readout principle at ESLAC. Only transitions responsible for the PDMR contrast between the different electron and nuclear spin states are visualized. The $|m_s\rangle = |1\rangle$ state, which was not probed in these experiments, is omitted for clarity. Under the application of high magnetic field (~510 G for ESLAC), in the ground state (**GS**) of the NV- centre the energy levels of $|m_s\rangle = |-1\rangle$ and $|m_s\rangle = |0\rangle$ are well separated (~1.5 GHz), whereas in the excited state (**ES**) they become nearly degenerate resulting in spin mixing between the states with equivalent total spin projection quantum number. The spin mixing combined with the electron spin polarization to $|m_s\rangle = |0\rangle$ through the metastable state (**MS**) [grey arrows], results in the spin polarization to the $|m_s\rangle = |0\rangle$ electron and $|m_I\rangle = |1\rangle$ nuclear spin state. The yellow arrows depict optical transitions induced by the application of the yellow-green laser. As can be seen, the $|m_s\rangle = |0\rangle$ spin sublevels in the ES are more likely to be excited by the second photon and contribute to the photocurrent by promoting the NV electron to the diamond conduction band (**CB**). When this happens, the negatively charged NV- centre is converted to NV0 centre (red arrows). The back-conversion is possible by another two-photon process while preserving the nuclear spin orientation. First, the NV0 centre is excited to the ES and subsequently, an electron is promoted from the valence band (**VB**) to the vacated orbital of NV0, leading to the formation of NV- centre. In this process the NV- $|m_s\rangle = |0\rangle$ ground states efficiently repolarise. **b**, Pulsed PDMR measurements of the NV nuclear (14N) and electron ($m_s$ = -1) spin hyperfine interaction for different magnetic fields showing nuclear spin polarization close to the ESLAC (for measurement at 123 G – 1500 ns laser pulse of 4 mW power, 1100 ns long MW π-pulse; for measurement at 439 G – 1000 ns laser pulse of 6 mW power, 400 ns long MW π-pulse).

3) Modelling

In order to provide a theoretical model describing the spin and photo-electric transitions, we model the NV centre system using the Lindblad master equation. The modelling allows us to calculate the resulting spin contrast and compare PMDR and optical detection of magnetic resonances (ODMR) readouts at various experimental conditions. We utilize the following formalism to calculate the time dependence of the density matrix

$$\dot{\rho} = -\frac{i}{\hbar}[H,\rho] + \sum_k \Gamma_k \left( L_k \rho L_k^\dagger - \frac{1}{2}\{L_k L_k^\dagger, \rho\} \right), \qquad (1)$$

where $H$ and $\rho$ are the Hamiltonian and the density matrix of the system, $L_k$ are Lindblad jump operators carrying out non-unitary transitions with rates $\Gamma_k$. The model includes five electronic states, namely the ground state, optical excited state, and the singlet shelving state in the negative charge state as well as the ground and excited states in the neutral charge state of the NV centre. In the simulations, we use 13 Lindblad operators that describe the transitions between the different electronic states. The spin Hamiltonian includes the electron, nuclear and hyperfine spin interactions in the triplet ground and excited state of the NV$^-$.

In order to determine power-dependent ionization rates, we record time traces of the optical signal starting from different initial spin states of the NV$^-$. These time traces reveal important details of the charge state dynamics of the NV centre and allow for validation of the model, which reproduces the experimental PL curves (see Figures 3a and b). In addition, it allows us to predict the time-dependent electron and hole currents. The time dependence of the normalized electron current closely follows the PL time traces (see Figures 3d and e). We attribute this similarity in the dynamics of the photon emission and photoionization to predominant contribution from the occupation of the NV$^-$ excited states for both. On the other hand, we note that the amplitudes of the spin-dependent PL and photocurrent signals show different power dependence. Besides electrons, emitted holes additionally contribute to the final spin contrast, prolonging it on longer timescales. Considering the power dependence of the contrast, we obtained larger ODMR contrast than PDMR contrast at 1 mW and comparable ODMR and PDMR contrasts at 3 mW (see Figures 3c and f). While this trend is similar to what has been observed in our experiments, our microscopic model on the isolated NV centre cannot quantitatively account for all the power-dependent observations. Therefore, we conclude that the residual deviation between the experimental and theoretical ODMR and PDMR contrasts are not related to the intrinsic properties of the NV centre but rather to its environment and further investigation into the charge state dynamics could be of interest. Charge carrier traps and recombination centres may significantly affect the electron and hole currents and their ratio, unlike emitted photons. Indeed, nonlinear recombination processes, that may be substantial at high laser powers, are able to enhance PDMR contrast. The fundamental differences in the detection of ODMR and PDMR signals may give rise to different contrasts that in turn enable further engineering of the PDMR contrast by material design and device fabrication.

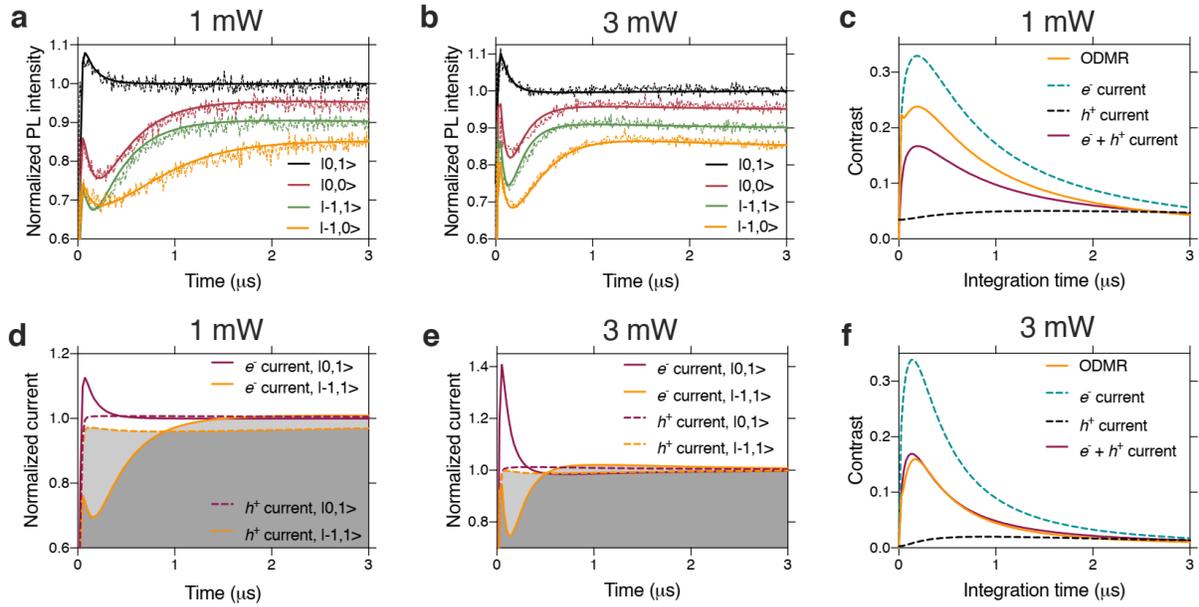

**Figure 3 | Photon and charge carrier emission dynamics and intrinsic spin contrast**. **a** and **b** depict the experimental (thin) and theoretical (bold) time-traces of the normalized PL intensity upon turning on the laser excitation pulse of 1 mW and 3 mW power, respectively, for various initial electron and nuclear spin states. For better visibility, the curves corresponding to |0,0⟩, |-1,1⟩, and |-1,0⟩ initial states are shifted down by 0.05, 0.1, and 0.15 respectively. To prepare different initial spin states in the experiment, RF, MW0, and MW1 π-pulses were used (see Figure 4a). The theoretical results closely follow the experimental curves. **d** and **e** depict the time traces of the simulated electron currents (solid lines) and hole currents (dashed lines) for 1 mW and 3 mW laser power, respectively, for |0,1⟩ and |-1,1⟩ initial spin states. The curves are normalized to the steady-state electron current obtained for the |0,1⟩ initial spin state. The areas below the electron and hole currents (grey areas under the solid and dashed lines) integrate to the same value to ensure the charge neutrality constraint of the photoionization cycle. The difference between the solid (dashed) curves provides the spin contrast of the electron (hole) current. **c** and **f** depict the theoretical ODMR contrast and the contrast of the electron-only current, the hole-only current, and the total current (PDMR) as a function of integration time for 1 mW and 3 mW excitation power, respectively. In the case of ODMR and PDMR, the contribution of the experimental background signal is taken into consideration as well.

4) Coherent electronic and nuclear qubit rotations

To coherently drive and read out the nuclear spin at the ESLAC, we used a set of RF pulses, applied to the nuclear spin, combined with MW assisted electron spin readout. First, using the relatively high nuclear spin polarization at 439 G, we probed the transition between |0,+1⟩ and |-1,0⟩ sublevels (See Figure 4a). To include the radiofrequency (RF) driving necessary to manipulate the nuclear spin, the envelope lock-in readout technique[3] was modified (see Figure 4b). The pulse sequence consisted of a laser pulse for spin polarization and photocurrent readout, followed by an RF pulse to drive the nuclear spin from |$m_I$⟩ = |+1⟩ state to |$m_I$⟩ = |0⟩ state and a 400 ns long MW π-pulse (MW0), set to selectively swap the |$m_I$⟩ = |0⟩ electron spin from |$m_s$⟩ = |0⟩ state to |$m_s$⟩ = |-1⟩ state (see Figure 4a). In this measurement, similarly to optical detection, the nuclear spin contrast is measured through the signal amplitudes of the particular spin states. To measure the resonance frequency of the nuclear spin, an RF pulse with a length corresponding to a π-rotation on the Bloch sphere was applied whilst sweeping its frequency (see Figure 4c). The laser power was kept at 6 mW with a pulse duration of 4 μs where single nuclear spin PDMR showed a contrast close to 5%. To demonstrate nuclear Rabi oscillations, the resonant RF pulse length was increased stepwise, while the distance between laser pulses was kept constant to conserve the same number of pulses per envelope duty cycle. The resulting electrically-detected Rabi nutations (see Figure 4d) maintained high contrast even for long RF pulse durations (up to 140 μs). We note that, depending on the duration of the nuclear spin rotation, these operations already correspond to a two-qubit CNOT gate, for which the electron and nucleus reach their highest level of entanglement at the π/2-point (~18 μs, in Figure 4). While we do not characterize their performance here, such gates can reach extremely high fidelity in NV centres, even at room temperature[14].

Unlike optical detection, where a short window (typically 300 ns) at the start of the readout pulse is recorded, photocurrent is integrated during the entire pulse sequence, as to-date current pre-amplifiers do not allow for sufficiently fast (>10 MHz) gating with high ($10^{12}$) current amplification at room temperature. Photocurrent integration therefore reduces the maximum detected spin contrast, as both the electron and nuclear spins are initialized within the first ~0.5 μs of the laser pulse[15]. Even though the detected contrast decreases using longer laser pulses and higher laser powers (see Figures 3c and f), due to intrinsic characteristics of the readout, PDMR contrast is maintained over prolonged time periods due to the hole current as devised from the theoretical modelling above (note that the optical curves in Figures 3a and b are shifted for clarity). A further improvement of the detected PDMR contrast at the prolonged measurement duration for the $|m_I\rangle = |-1\rangle$ and $|m_I\rangle = |0\rangle$ nuclear spin state polarisation is given by the spin mixing at ESLAC, which spans 3× and 2× longer than the $|m_I\rangle = |+1\rangle$ state[12]. Furthermore, by lowering the technical noise thanks to the small electrode architecture and lock-in readout technique a sufficiently high detected nuclear PDMR contrast was reached in our measurement. Development of fast current preamplifiers will enable further significant improvement of the measured spin contrast.

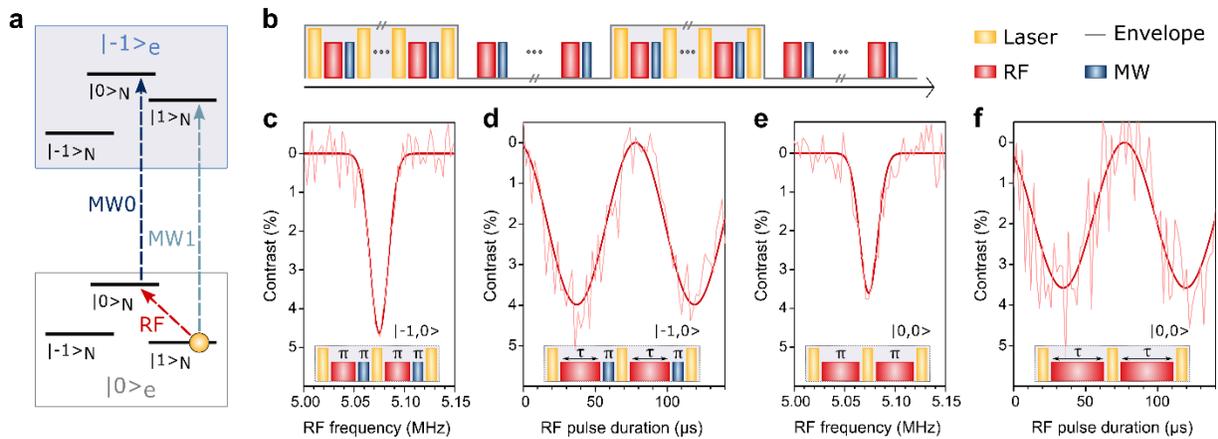

**Figure 4 | Electrical readout of individual nuclear spin. a,** Schematics of the NV ground state hyperfine energy-level structure of the $|m_s\rangle = |0\rangle$ and $|m_s\rangle = |-1\rangle$ states depicting the resonant frequencies of the probed

transitions. (RF – radiofrequency, MW0 – resonant microwave frequency to selectively excite electron spin in $|m_I\rangle = |0\rangle$), MW1 – resonant microwave frequency to selectively excite electron spin in $|m_I\rangle = |+1\rangle$). **b**, Scheme of the envelope pulse train designed for electrical readout of single nuclear spin using lock-in detection technique. Here, the lock-in amplifier is triggered by the on/off envelope modulation of the laser pulses. **c**, Electrically-detected RF resonant frequency of the $|0,+1\rangle$ and $|0,0\rangle$ transitions of the single $^{14}$N nuclear spin measured at 439 G. Inset shows the pulse sequence used, consisting of RF and MW0 π-pulses. **d**, Corresponding electrically-detected Rabi oscillations of the single nuclear spin with the pulse sequence shown in the inset. **e**, Electrically-detected RF resonant frequency of the $|0,+1\rangle$ and $|0,0\rangle$ transitions of the single $^{14}$N nuclear spin without electron spin manipulation measured at 439 G. Inset shows the pulse sequence used consisting of RF π-pulses. **f**, Corresponding electrically-detected Rabi oscillations of the single nuclear spin with the pulse sequence shown in the inset. (Experimental conditions: 4000 ns laser pulse of 6 mW power, 400 ns long MW π pulse, 1 W RF power).

The scheme presented above is realised by both the MW (electron) and RF (nuclear) spin driving. However, in many applications it might be advantageous to use MW-free readout (e.g. to reduce additional heating in MW absorbing environment and/or to avoid the MW resonant frequency jitter in long measurements). Interestingly, the spin mixing between the nuclear and electron spins at the ESLAC permits to achieve such MW-free readout by the reverse polarisation of the electron spin by the nuclear spin. To demonstrate this possibility, we modified the detection sequence excluding the MW driving pulses. Figure 4e and Figure 4f show nuclear resonance and a Rabi nutation's, respectively, for the MW-free situation. Surprisingly, we achieved a contrast similar to the MW-assisted readout using the MW-free detection, indicating that the electron spin can be repolarised by the nuclear spin very efficiently, leading to low contrast losses. The measurements depicted in Figure 4c-f were additionally repeated at higher B-field of 500 G yielding similar results.

## Conclusion

The experimental results presented in this paper are the first demonstration of electrical detection of a single nuclear spin at room-temperature, achieved using two-qubit spin gates applied to the electron spin and the intrinsic $^{14}$N nuclear spin of the NV centre. The developed theoretical model allows for a precise description of spin-dependent photocurrent and its time dependency at excited-state level anti-crossing (ESLAC) and can be used as a highly efficient spin control and readout technique. It shows that PDMR keeps a high spin contrast even for prolonged sequences used for nuclear spin manipulation. Electrical readout can now be applied to far more complex gate sequences, with the benefits of long nuclear spin coherence, opening the possibility for technologically simpler realizations of small quantum processors. Downscaling the size of the electrodes provides a prospect towards the development of site-resolved readout of dipole-dipole entangled NV qubits, and thereby towards the realization of nanoscale coupled registers, in future works. The facile integration of PDMR with classical electronics, thus opens new perspectives for the further development of scalable diamond quantum devices at room-temperature.

# Acknowledgements

M.N. acknowledges the QuantERA projects Q-Magine and NanoSense funded through the Flemish Scientific Foundation (FWO) as well as FWO projects G0E7417N and G0A0520N and Quantum Flagship project ASTERIQS, No 820394. M.N and M.G also acknowledge the project No. GA20-28980S funded by the Grant Agency of the Czech Republic. D.W. and M.T. acknowledge the support of the FFG through projects 864036 (QuantERA Q-Magine) and 870002 (QSense4Life). V.I. acknowledges the support from the MTA Premium Postdoctoral Research Program and the Knut and Alice Wallenberg Foundation through WBSQD2 project (Grant No. 2018.0071). V.I. and A.G. acknowledge the Hungarian NKFIH grants No. KKP129866 of the National Excellence Program of Quantum-coherent materials project and No. NN127889 of the EU QuantERA program Q-Magine. VI and AG acknowledge support of the NKFIH through the National Quantum Technology Program (Grant No. 2017-1.2.1-NKP-2017-00001) and the Quantum Information National Laboratory sponsored by Ministry of Innovation and Technology of Hungary. The calculations were performed on resources provided by the Swedish National Infrastructure for Computing (SNIC 2018/3-625 and SNIC 2019/1-11) at the National Supercomputer Centre (NSC). J.H. is a PhD fellow of the Research Foundation - Flanders (FWO).